\documentclass[onecolumn,useAMS]{mn2e}
\usepackage{epsfig}
\usepackage{times} 
\def\apj{ApJ}
\def\mnras{MNRAS}
\def\apjl{ApJL}
\def\nat{Nature}
\def\aj{AJ}
\def\aap{A \& A}
\def\la{\lower.5ex\hbox{$\; \buildrel < \over \sim \;$}}
\def\ga{\lower.5ex\hbox{$\; \buildrel > \over \sim \;$}}
\def\physrep{Physics Reports}

\begin{document}

\title[Large Scale Magnetic Fields:  Galaxy Two-Point correlation function]{Large Scale Magnetic Fields:  Galaxy Two-Point correlation function}
\author[Shiv K. Sethi]
{Shiv K. Sethi \\
\hspace{-0.1cm}Raman Research Institute, Bangalore 560080, India \\
\hspace{-0.1cm} email: sethi@rri.res.in
}

\maketitle
\begin{abstract}
We study the effect of large scale tangled magnetic fields on the galaxy
two-point correlation function in the redshift space. We show that (a) the 
magnetic field effects can be comparable the gravity-induced clustering for
present magnetic field strength $B_0 \simeq 5 \times 10^{-8} \, \rm G$,  (b)
the absence of this signal from the present data gives an upper bound
 $B_0 \la 3 \times 10^{-8} \, \rm G$, (c) the future data can probe the 
magnetic fields of $\simeq 10^{-8} \, \rm G$. A comparison with other constraints on the present magnetic field shows that they are marginally compatible.
 However  if the magenetic fields corresponding 
to $B_0 \simeq 10^{-8} \, \rm G$ 
existed at
the last scattering surface  they will cause unacceptably large CMBR anisotropies.

\smallskip

\end{abstract}

\section{Introduction}
Spatially coherent magnetic fields
 are ubiquitous in galaxies and galaxies clusters
(for a recent review see Widrow 2002). It is not well understood 
whether these fields are flux-frozen primordial fields or they originated
by the dynamo amplification of small seed fields.  
The existence of  magnetic fields at larger scales 
($\ga 1 \,\rm Mpc$)
cannot yet be inferred from direct observations (for a summary of 
results Kronberg 1994,  Widrow 2002). 
If these large scale primordial fields exist they
can serve, depending on their strength, either of these two scenarios.  
Therefore  their existence might   be of great importance to
understand the way galaxies and clusters formed.  Also they 
can be dynamically important in shaping the 
large scale structure in the universe (Wasserman 1978).

Wasserman (1978) showed that the large scale magnetic fields can generate
density and velocity perturbations which could be responsible for the 
formation of structures in the universe. Kim, Olinto \& Rosner (1996)
studied this hypothesis in more detail. In this paper, we attempt to
understand the effect of primordial magnetic fields on the galaxy
two-point correlation function.

One of the most important diagnostic of   structures formation in the universe
 is the two-point correlation function of the galaxy distribution (Peebles 1980). Recently the two-point correlation function in the redshift space was accurately determined  at large scales ($\ga 10 \, \rm Mpc$) (Peacock {\it et al.} 2001). This gives the best
statistical evidence of the large scale velocity flows in the universe. 
Analysis shows this velocity flow to be consistent with the assumption
that the structures in the universe formed from  gravitational
instability (Hawkins {\it et al.} 2002, Peacock {\it et al.} 2001).
 The present and up-coming galaxy surveys like 2dF
(Colless {\it et al.} 2001) and SDSS (York {\it et al.} 2000) 
 are  large enough  to do precision cosmology. Therefore it is of interest to
ask whether galaxy distribution at the present epoch could also be 
affected by causes other than gravitational instability. 

In this paper we estimate the  two-point 
correlation function in redshift space from primordial tangled 
magnetic field. The velocity flow in the presence of magnetic fields
will not be pure gradient as is the case of gravity. Such velocity 
flows might leave unique geometrical signals in the correlation function.
On the other hand lack of such signatures in the data can put meaningful
upper bounds on the large scale magnetic fields. 
  
The most direct method to infer the presence of intergalactic magnetic 
fields for $z \la 3$ is to study the Faraday rotation of polarized emission from
extra-galactic sources (Rees \& Reinhardt 1972, Kronberg \& Simard-Normandin 1976, Vall\'ee 1990, Kolatt 1998, Blasi, Burles \& Olinto 1999). The existence
of these  fields 
can also be constrained at the last scattering surface from CMBR anisotropy and 
spectral distortion measurements (Barrow, Ferreira \& Silk 1997,  Subramanian \& Barrow 1998, Jedamzik, Katalini{\' c}, \& Olinto  2000) and Faraday rotation of the 
polarized component of the CMBR anisotropies (Kosowsky \& Loeb 1996). If magnetic fields existed at even higher redshifts, they can also affect the primordial
nucleosynthesis (see e.g. Widrow 2002 for detailed discussion).

In the next section we briefly discuss the equations of motion of 
the cosmological fluid in the presence of magnetic fields and other 
preliminaries for this study. In \S 3 we calculate the two-point correlation
function in the redshift space including the effect of magnetic fields.
In \S 4 we present  our results and compare them with other
constraints on the present magnetic fields. In \S 5 we give our 
conclusions. Throughout this paper we 
use the currently-favoured background cosmological model: spatially flat
with $\Omega_m = 0.3$ and $\Omega_\Lambda = 0.7$ (Perlmutter {\it et al.} 1999,
Riess {\it et al.} 1998). For numerical work we use $\Omega_b h^2 = 0.02$ (Tytler {\it et al.} 2000) and 
$h = 0.7$ (Freedman {\it et al.} 2001). 

\section{Magneto-hydrodynamics Equations }
In linearized Newtonian theory  the equations of magneto-hydrodynamics for one 
fluid model \footnote{in this model all the three components of the matter, electrons, protons
and the neutral hydrogen particles share the force applied on the
small ionized component in the post-recombination era and the small 
ionized component prevents the magnetic fields from decaying.} in the 
co-moving coordinates   are (Wasserman 1978):
\begin{eqnarray}
{d (a {\bf  v_{\rm \scriptscriptstyle b}}) \over dt}  & = & - {\bf \nabla} \phi + {({\bf \nabla} \times {\bf B}) \times  {\bf B} \over 4 \pi \rho_{\scriptscriptstyle b}} \label{eq:n1}\\
\nabla .{\bf v_{\rm \scriptscriptstyle b}} & = & -a \dot \delta_{\scriptscriptstyle b}  \label{eq:n2} \\
\nabla^2 \phi & = & 4 \pi G  a^2 (\rho_{\scriptscriptstyle DM}\delta_ {\scriptscriptstyle DM} + \rho_{\scriptscriptstyle b} \delta_{\scriptscriptstyle b})\label{eq:n3p}\\
{\partial (a^2 {\bf B}) \over \partial t}& = &{\nabla \times ({\bf v_{\scriptscriptstyle b}} \times a^2 {\bf B}) \over a} \label{eq:n3}\\
\nabla . {\bf B} & = &  0 \label{eq:n4}
\end{eqnarray}
In Eq.~(\ref{eq:n1}) we have neglected the pressure gradient term on the 
right hand side as it is important  at Jeans' length scales ($\ll 1 \, \rm Mpc$
before re-ionization and $\simeq 1 \, \rm Mpc$ after re-ionization). Our
interest here is to study 
scales at which the perturbations are linear at the present epoch, $\ga 10 \, \rm h^{-1} Mpc$. 
Eq.~(\ref{eq:n1}) and Eq.~(\ref{eq:n2}) can be combined to give:
\begin{equation}
{\partial^2\delta_{\scriptscriptstyle b} \over \partial t^2} +2{\dot a \over a} {\partial \delta_{\scriptscriptstyle b} \over \partial t} - 4 \pi G   (\rho_{\scriptscriptstyle DM}\delta_ {\scriptscriptstyle DM} + \rho_{\scriptscriptstyle b} \delta_{\scriptscriptstyle b}) =-{{\bf \nabla .}\left [({\bf \nabla} \times {\bf B}) \times  {\bf B} \right ] \over 4 \pi a^2 \rho_{\scriptscriptstyle b}}
\label{eq:n4p}
\end{equation}
Here the subscript 'b' refers to the baryonic component and the 
subscript 'DM' refers to  the dark matter. 
We do not give here the evolution of the 
dark matter perturbations which is well known (Peebles 1980) and 
 can be solved using the equations above 
by dropping the magnetic field terms. It was shown by Wasserman (1978) that Eq.~(\ref{eq:n4p}) admits a growing 
solution, i.e. tangled magnetic fields can lead to growth the density
perturbation. These solutions are discussed in Appendix A. 
In writing Eq.~(\ref{eq:n3}) we have assumed the medium to
have infinite conductivity. It  can be simplified further by dropping the 
right hand side of the equation as it is of higher order, this gives:
\begin{equation}
{\bf B}(x,t)a^2 = \hbox{constant}.
\label{eq:n5}
\end{equation}

We assume the tangled magnetic field to be a statistically homogeneous and 
isotropic vector  random process. In this case the two-point correlation
function of the field in Fourier space can be expressed as (Landau \& Lifshitz 1987):
\begin{equation}
\langle {B_i({\bf q})B^*_j({\bf k})} \rangle = \delta^3_{\scriptscriptstyle D}({\bf q -k}) \left (\delta_{ij} - q_i q_j/q^2 \right ) B^2(q)
\label{eq:n6}
\end{equation}

\section{Two-point correlation function in Redshift Space}

The density field in the universe is a statistically homogeneous and 
isotropic random process in the real space. 
However in redshift space  it becomes both 
statistically inhomogeneous and anisotropic (for a detailed discussion on this
point see Hamilton 1998 and reference therein). The measured position of 
an object, {\bf s}, is related to the real position, {\bf r} at the 
present redshift  as:
\begin{equation}
{\bf s}  = {\bf r} + H_0^{-1} {\bf (\hat r. v_{\rm \scriptscriptstyle b}) \hat r}
\label{eq:n7}
\end{equation}
In linearized theory the density inhomogeneity in redshift space is related
to the density inhomogeneity in the redshift space as:
\begin{equation}
\delta^s({\bf r}) = \delta({\bf r}) -({\bf \hat r. \nabla} + \alpha({\bf r})/r){\bf \hat r.v_{\rm \scriptscriptstyle b}}
\label{eq:n8}
\end{equation}
Here $\alpha(r) = \partial \ln r^2 \bar n(r)/\partial \ln r$ is the logarithmic
derivative of the survey selection function $\bar n(r)$. Galaxy correlation
function  is expected to be biased with respect to the underlying density field  correlations (Bardeen {\it et al.} 1986). Recent results show that optical
galaxies are unbiased tracers of the underlying density field at linear
scales ($\ga 10 \, \rm Mpc$) (Lahav {\it et al.} 2002, Verde {\it et al.} 2002).
We take the value of linear bias to be  one throughout this paper. 

We use the plane parallel approximation in this paper (Kaiser 1987). 
In this approximation 
the term containing $\alpha(r)$ in Eq.~(\ref{eq:n8}) can be dropped and lines
of sight to two different sources on the sky can be taken to be parallel. 
This is valid  either if the sources are far away and/or the angle between the 
two lines of sight $\theta \ll 1$ (see Hamilton 1998 for details). 
The two-point correlation function in
redshift space can then be written as:
\begin{equation}
\langle \delta^s({\bf r'}) \delta^s({\bf r'+r}) \rangle = \left\langle (\delta({\bf r'}) -  {\bf \hat z. \nabla}{\bf \hat z.v_{\rm \scriptscriptstyle B}({\bf r})})  (\delta({\bf r +r' }) -  {\bf \hat z. \nabla}{\bf \hat z.v_{\rm \scriptscriptstyle B}({\bf r+r'})}) \right \rangle
\label{eq:redsd}
\end{equation}
Here ${\bf \hat z}$ is the unit vector in the z direction, which is also 
chosen to be the line of sight. The density contrast and velocity perturbation
in Eq.~(\ref{eq:redsd}) get contribution from both the magnetic field
induced perturbations and gravity-induced perturbations (the solution of 
the homogeneous part of Eq.~(\ref{eq:n4p})).    
We assume the 'initial condition'  perturbations to
be uncorrelated with the magnetic field induced perturbations. This 
is justified as the only coupling between these components comes from the 
potential fluctuation term (Eq.~(\ref{eq:n3p})). As $\rho_b \ll \rho_{\scriptscriptstyle DM}$ (Peacock {\it et al.} 2001, Tytler {\it et al.} 2000), this 
correlation can be neglected or in other 
words we assume the baryons to contribute negligibly
to the potential fluctuations. 
   These simplifications  allow us to 
separate the magnetic field induced density and velocity perturbations.
For the rest of this paper we only deal with magnetic field  induced perturbations
and  $\delta_b$ and $v_b$ refer to these 
perturbations. 

The time dependent part of Eqs~(\ref{eq:n1}) and~(\ref{eq:n4p}) can be solved
to give (Appendix A):
\begin{eqnarray}
\delta_{\scriptscriptstyle b}({\bf r}, t) & = &  g(t) \, {\bf \nabla .}\left [({\bf \nabla} \times {\bf B}) \times  {\bf B} \right ] \\
{\bf v_{\rm \scriptscriptstyle b}}({\bf r},t) & = & q(t) \,({\bf \nabla} \times {\bf B}) \times  {\bf B}
\end{eqnarray}

 Eq.~(\ref{eq:redsd}) has three terms: the density-density correlation, the density-velocity
correlations and the velocity-velocity correlation. For fluctuations seeded 
by magnetic fields the  three terms in Eq.~(\ref{eq:redsd}) can be 
expanded, using Eq.~(\ref{eq:n6}),  as:
\begin{eqnarray}
\langle \delta _{\rm \scriptscriptstyle b}({\bf r'}) \delta_{\rm \scriptscriptstyle b}({\bf r +r' }) \rangle = & &\! \! \! \! \! \! \! \!\! \! \! \!  \! g^2(t)\int {d^3k_1 \over (2 \pi)^3} {d^3k_2 \over (2 \pi)^3} \exp  \left [({\bf k_1} + {\bf k_2}){\bf . r}\right ] B^2(k_1) B^2(k_2) \times \nonumber \\ \biggl [ k_1^4 + 3k_1^2 k_2^2   
    + ({\bf k_1.k_2}) (k_2^2  - k_1^2)  \!\! & +  & \! \! ({\bf k_1.k_2})^2 \left ({k_1^2 \over k_2^2} -3 \right ) + {({\bf k_1.k_2})^3 \over k_1^2 k_2^2}(k_2^2-k_1^2) \biggr]
\label{eq:expa1}
\end{eqnarray}
\begin{eqnarray}
\left \langle \delta_{\rm \scriptscriptstyle b}({\bf r'}){\bf \hat z. \nabla}{\bf \hat z.v_{\rm \scriptscriptstyle b}}({\bf r+r'})  \right \rangle   = &   & \! \! \! \! \!\! \! \! \! \!g(t) q(t)\int {d^3k_1 \over (2 \pi)^3} {d^3k_2 \over (2 \pi)^3} \exp  \left [({\bf k_1} + {\bf k_2}){\bf . r}\right ] B^2(k_1) B^2(k_2) \times \nonumber \\ 
\Biggl [ k_{\rm 1z}^2(3k_2^2+k_1^2) +(k_1^2 + k_2^2)k_{\rm 1z}k_{\rm 2z} \!\!  &+ & \!\! 
  ({\bf k_1.k_2})\left( 5k_{\rm 1z}^2 +
{k_{\rm 1z}^2 k_2^2 \over k_1^2}+3 k_{\rm 1z}k_{\rm 2z}\right)  \nonumber \\
   +({\bf k_1.k_2})^2 \left ({k_{\rm 1z}^2 \over k_2^2}+{k_{\rm 1z}^2 \over k_1^2} - {k_{\rm 1z}k_{\rm 2z} \over k_2^2} \right ) \Biggr ] &&
\label{eq:expa2}
\end{eqnarray}
Here $k_z = {\bf k . \hat z}$ is the angle between the k-mode and the line 
of sight. 
\begin{eqnarray}
\left \langle {\bf \hat z. \nabla}{\bf \hat z.v_{\rm \scriptscriptstyle b}({\bf r})} {\bf \hat z. \nabla}{\bf \hat z.v_{\rm \scriptscriptstyle b}({\bf r + r'})} \right \rangle & = & \!\! \! \! \! q^2(t) \int {d^3k_1 \over (2 \pi)^3} {d^3k_2 \over (2 \pi)^3} \exp  \left [({\bf k_1} + {\bf k_2}){\bf . r}\right ] B^2(k_1) B^2(k_2)\times \nonumber \\
\biggl [ \left (k_{\rm 1z}^4 + 3k_{\rm 1z}^2 k_{\rm 2z}^2 + 3k_{\rm 1z}^3k_{\rm 2z}+ 
k_{\rm 1z}k_{\rm 2z}^3 \right . \! \! & + & \! \! \left .  k_{\rm 1z}^2 k_2^2 \right )-({\bf k_1.k_2})\left ({4k_{\rm 1z}^3k_{\rm 2z} \over k_1^2} + {2 k_{\rm 2z}^3 k_{\rm 1z} \over k_2^2} +
{k_{\rm 1z}^2 k_{\rm 2z}^2 \over k_2^2} \right . \nonumber \\
\left .+ {k_{\rm 1z}^2 k_{\rm 2z}^2 \over k_1^2} +{2 k_{\rm 1z}^4 \over k_1^2} \right ) +    {({\bf k_1.k_2})^2 \over k_1^2 k_2^2}\! \!\! \! \! \!&  &\! \!\! \! \! \!\left (k_{\rm 1z}^4 + 9k_{\rm 1z}^2 k_{\rm 2z}^2 + 6k_{\rm 1z}^3k_{\rm 2z}+ 2k_{\rm 2z}^3k_{\rm 1z}- k_{\rm 1z}^2 k_2^2 \right ) \biggr ]
\label{eq:expa3}
\end{eqnarray}

For studying fluctuations in redshift space, quantities of interest are 
the angular moments of the redshift space two-point correlation function along the line of sight (Hamilton 1992):
\begin{equation}
\xi_\ell({\bf r}) = \int d\mu_r \, P_\ell(\mu_r) \langle \delta^s({\bf r'}) \delta^s({\bf r'+r}) \rangle
\label{eq:momleg}
\end{equation}
Here $\mu_r = {\bf r . \hat z}$ is the angle between the separation between
the two points and the line of sight and $P_\ell(\mu_r)$ are Legendre polynomials. Integrating Eqs~(\ref{eq:expa1})~(\ref{eq:expa2}), and~(\ref{eq:expa3})
 over angles and  writing in terms of moments about the 
line of sight, we get:
\begin{eqnarray}
\langle \delta_{\rm \scriptscriptstyle b}({\bf r'}) \delta_{\rm \scriptscriptstyle b}({\bf r +r' }) \rangle  \!  = & & \! \! \! \! \! \!  g^2(t)   \int {k_1^2dk_1 \over 2 \pi^2} {k_2^2 dk_2 \over 2 \pi^2}  B^2(k_1) B^2(k_2)
 P_0 (\mu_r)\times \nonumber \\
\biggl [j_0(k_1 r)j_0(k_2 r) \left({2 \over 3} k_1^4+2k_1^2 k_2^2 \right ) & + &
j_2(k_1 r) j_2(k_2 r) \left ({2 \over 3} k_1^4-2k_1^2 k_2^2 \right ) \biggr ]
\label{eq:isotro}
\end{eqnarray}

\begin{eqnarray}
\left \langle \delta_{\rm \scriptscriptstyle b}({\bf r'}){\bf \hat z. \nabla}{\bf \hat z.v_{\rm \scriptscriptstyle b}}({\bf r+r'})  \right \rangle   = & & \! \! \! \! \! \! \! \!\! g(t)q(t)\int {k_1^2 dk_1 \over 2 \pi^2} {k_2^2 dk_2 \over 2 \pi^2}  B^2(k_1) B^2(k_2) \times \nonumber \\
\biggl \{P_0(\mu_r) \biggl[ j_0(k_1 r)j_0(k_2 r) \left({1 \over 9} k_1^4+{4 \over 9}k_1^2 k_2^2 \right ) \!\! & + & \!\! j_1(k_1 r)j_1(k_2 r) \left({31 \over 15} k_1^3 k_2 \right) + \nonumber \\
 j_2(k_1 r)j_2(k_2 r) \left({1 \over 9} k_1^4+{7 \over 9}k_1^2 k_2^2 \right ) \!\! & + & \!\! {1\over 5} k_1^3 k_2 \left (j_3(k_1 r)j_1(k_2 r) + j_1(k_1 r)j_3(k_2 r) \right) +\nonumber \\
 {2 \over 15} k_1^3k_2 j_3(k_1 r)j_3(k_2 r) \biggr ] \! & + & \! \! \! \!P_2(\mu_r) \biggl [ j_2(k_1 r)j_0(k_2 r) \left (-{2 \over 3}k_1^3k_2-2k_2^3k_1 -{4 \over 3} k_1^2k_2^2 \right ) \nonumber \\
-{229 \over 75}k_1^3k_2 j_1(k_1 r)j_1(k_2 r)\!\! &+&\! \! j_2(k_1 r)j_2(k_2 r)\left({50 \over 63}k_1^2k_2^2 + {8 \over 63} k_1^4 \right )\nonumber \\
 & - & j_4(k_1 r)j_2(k_2 r)\left ({8 \over 35} k_1^2k_2^2 + {8 \over 35} k_1^4 \right ) \biggr ] \biggr \} 
\label{eq:velden}
\end{eqnarray} 
\begin{eqnarray}
\left \langle {\bf \hat z. \nabla}{\bf \hat z.v_{\rm \scriptscriptstyle b}({\bf r})} {\bf \hat z. \nabla}{\bf \hat z.v_{\rm \scriptscriptstyle b}({\bf r + r'})} \right \rangle = & & \! \! \! \!  q^2(t)\int {k_1^2dk_1 \over 2 \pi^2} {k_2^2dk_2 \over 2 \pi^2}  B^2(k_1) B^2(k_2)\times \nonumber \\
\biggl \{P_0(\mu_r) \biggl[ j_0(k_1 r)j_0(k_2 r) \left({4 \over 15} k_1^4-{46 \over 225}k_1^2 k_2^2 \right ) \!\! & + & \!\! j_1(k_1 r)j_1(k_2 r) \left({28 \over 15} k_1^3 k_2 \right) + \nonumber \\
 j_2(k_1 r)j_2(k_2 r) \left({1082 \over 7875} k_1^4+{1451\over 1575}k_1^2 k_2^2 \right ) \!\! & + & \! \! j_3(k_1 r)j_1(k_2 r) \left (-{466 \over 525}k_1^3k_2-{13 \over 225}k_1k_2^3 \right ) \nonumber \\
 +j_1(k_1 r)j_3(k_2 r) \left ({107 \over 175}k_1^3k_2-{58 \over 225}k_1k_2^3 \right )\!\! & + & \! \!j_3(k_1 r)j_3(k_2 r) \left (-{328 \over 1225}k_1^3k_2-{704 \over 11025}k_1k_2^3 \right ) \nonumber \\
 + {48 \over 175} k_1^2k_2^2 j_4(k_1 r)j_4(k_2 r) \biggr ] & + &  P_2(\mu_r) \biggl [j_2(k_1 r)j_0(k_2 r) \left (-{16 \over 21}k_1^4 -{7\over 15} k_1^2k_2^2 \right ) \nonumber \\
+ j_0(k_1 r)j_2(k_2 r) \left (-{8 \over 105}k_1^4 +{148\over 105} k_1^2k_2^2 \right )& - &j_1(k_1 r)j_1(k_2 r) \left({282 \over 75} k_1^3 k_2 \right) \nonumber \\
j_2(k_1 r)j_2(k_2 r) \left({1136 \over 11025} k_1^4+{470\over 1029}k_1^2 k_2^2 \right ) \!\!  & +  & \! \! j_3(k_1 r)j_1(k_2 r) \left (-{312 \over 245}k_1^3k_2-{208 \over 1225}k_1k_2^3 \right ) \nonumber \\
+ j_1(k_1 r)j_3(k_2 r) \left (-{288 \over 1225}k_1^3k_2-{632 \over 1575}k_1k_2^3 \right ) & - &  {792 \over 1225}k_1^3k_2 j_3(k_1 r)j_3(k_2 r) \nonumber \\
+ j_4(k_1 r)j_2(k_2 r) \left(-{1504 \over 8085} k_1^4+{32\over 1715}k_1^2 k_2^2 \right )\!\! \! & +  & \! \! \!{96 \over 343} k_1^2k_2^2 j_4(k_1 r)j_4(k_2 r) +{64 \over 147} k_1^3k_2 j_5(k_1 r)j_3(k_2 r) \biggr ] \nonumber \\
+ P_4(\mu_r) \biggl [ j_4(k_1 r)j_0(k_2 r) \left ({ 32\over 105}k_1^4 +{96\over 175} k_1^2k_2^2 \right ) \! \! & + & \! \!
j_2(k_1 r)j_2(k_2 r) \left({832 \over 18375} k_1^4+{1536\over 1715}k_1^2 k_2^2 \right ) \nonumber \\
j_3(k_1 r)j_1(k_2 r) \left ({1336 \over 1225}k_1^3k_2+{184 \over 1225}k_1k_2^3 \right )\! \! & + & \! \! j_1(k_1 r)j_3(k_2 r) \left ({176 \over 1225}k_1^3k_2-{488 \over 1575}k_1k_2^3 \right ) \nonumber \\
- {48 \over 539}k_1^3k_2 j_3(k_1 r)j_3(k_1 r) & - &
j_4(k_1 r)j_2(k_2 r)\left ({512 \over 8085} k_1^4  + {352 \over 1029} k_1^2 k_2^2 \right ) \nonumber \\
- {48 \over 315} k_1^3k_2 j_5(k_1 r)j_1(k_2 r) & + &  {1296 \over 8575} k_1^2k_2^2 j_4(k_1 r)j_4(k_2 r)  \nonumber \\ 
+ {128 \over 735} k_1^3k_2 j_5(k_1 r)j_3(k_2 r)
& + & {70128 \over 473473} k_1^4 j_6(k_1 r)j_2(k_2 r) \biggr ] \biggr \} 
\label{eq:velvel}
\end{eqnarray}

We normalize the  RMS of magnetic field which  can be written as:
\begin{equation}
B_0^2 \equiv \langle B_i({\bf x}) B_i({\bf x}) \rangle = {1 \over \pi^2} \int dk k^2 B^2(k)
\end{equation}
The magnetic field power spectrum  is taken to be power law:
\begin{equation}
 B^2(k) = A k^n
\end{equation}
 In normalizing the power spectrum we use a sharp-k filter
  with $k_c = 1 \, h \, \rm Mpc^{-1}$ (Subramanian \& Barrow 2002). This gives:
\begin{equation}
A = {\pi^2 (3+n) \over k_c^{(3+n)}} B_0^2 
\end{equation}

\section{Results}

Our main results are given in Eqs~(\ref{eq:isotro}), (\ref{eq:velden}) and
(\ref{eq:velvel}). The quantities of interest, as they are directly
measured by observations, are the line-of-sight angular moments of redshift-space correlations (Eq.~(\ref{eq:momleg})). As in the case 
of purely gravitational clustering, the three non-vanishing moments 
are the zeroth, second, and the forth; these are 
the coefficient of $P_0(\mu_r)$, 
$P_2(\mu_r)$, and $P_4(\mu_r)$, respectively. However the 'geometry' of 
magnetic field induced perturbations is more complicated.
We collect appropriate terms from Eqs~(\ref{eq:isotro}), (\ref{eq:velden}) and
(\ref{eq:velvel}) to calculate the three moments (Eq.~(\ref{eq:momleg})).

In Figure~1 we plot the three moments of correlation function for 
$B_0 = 5 \times 10^{-8} \, \rm G$ and magnetic spectral index $n = 1$ at the 
present epoch.
 For comparison the gravity induced 
moments are also plotted for the currently favoured value of $\beta = 0.43$ 
(Peacock {\it et al.} 2001). Note that only scales above $10 \, \rm h^{-1} \, Mpc$ are plotted, even though magnetic fields effects are larger
at smaller scales. It is because  non-linear clustering
effects become important at smaller scales. In
redshift space the effects of non-linear clustering become dominant at larger
scales as compared to the real space (for discussion and references see 
Hamilton 1998). At scales smaller than $\simeq 10 \, \rm h^{-1} \, Mpc$, the random velocities wash out the information contained in
linear coherent flows. In gravitational clustering scenario the seocnd moment of
the correlation function 
is driven to zero and changes sign below these scales (the finger-of-god effect; see Peacock {\it et al.} 2001 for recent observational evidence of 
cross-over from coherent flow to random flow). As we anticipate the 
overall velocity flows to be dominated by gravitational effects, in keeping with the 
good observational evidence in its support (Peacock {\it et al} 2001, Hawkins {\it et al} 2002) , it is difficult to get
information about magnetic field induced coherent flows at scales smaller
than $\simeq 10 \, \rm h^{-1} \, Mpc$ from data.

Can the present data show the evidence of large scales magnetic fields? 
Currently the largest redshift survey is the 2dF galaxy survey (Colless {\it et al. 2001}). Its
most recent analyses contain nearly 2{,}20000 galaxies (Hawkins {\it et al.} 2002). Two kind of statistics commonly used to infer large
scale coherent flows  are the angular-averaged  two-point
function (the zeroth moment) and the ratio of the second and the zeroth moment
(Hawkins {\it et al.} 2002). 
If magnetic field is present with a power law power spectrum then its
presence will show itself as oscillations in the two-point functions. 
The present data does not show such oscillations, so it can be used to 
put upper bound on the strength of the magnetic fields. It should however 
be borne in mind that interpretation of such statistics is difficult as
the incoherent velocity
flow continue to be important, though sub-dominant, at larger scales also
(see e.g. Hatton \&  Cole 1998; Peacock {\it et al} 2001). 
 To distinguish the effect of magnetic 
field induced flows from uncertainty owing to incoherent flows a typical
correlation function resolution of $\simeq 0.1$ would be required for 
scales $\ga 10 \, \rm h^{-1} \,  Mpc$  (Peacock {\it et al.} 2001). 
This already rules out the case shown in Figure~1.  
The current data constrains the magnetic field strength to be $B_0 \la 3 \times 10^{-8} \, \rm G$. A better way to detect magnetic field effects or put
upper limit on the magnetic field strength might be to extract the forth
moment of the correlation function. As seen in Figure~1 magnetic field
effect dominate over the pure gravity contribution for even a smaller
value of magnetic field strength. However it is noisier to extract the 
forth moment (Hamilton 1998) and it has not so far been computed from
the galaxy data. 

From future data like SDSS galaxy survey (York {\it et al.} 2000) it might be possible to detect
the effects of even smaller magnetic fields on two-point correlation function.
Notwithstanding the effects of incoherent flows and other
systematics, the theoretical limit on
the error of two point correlation function can be calculated. Landy and 
Szalay (1993) showed that by a judicious choice of two-point 
correlation function estimator,  error on it 
 is given by the Poisson noise, i.e. 
$\Delta \xi = 1/n_p(r)^{1/2}$; $n_p(r)$ is the number of galaxy pairs for 
separation between $r$ and $r+dr$. It can be written as $n_p(r) = f(r) n(n-1)/2$, 
where $n$ is the 
total number of objects in the sample and $f(r)$ is the fraction of 
pairs at a separation $r$. Taking $n = 10^{6}$ from SDSS galaxy survey and 
$f(r)= 10^{-4}$ for some scale of interest, $ \Delta \xi \simeq 10^{-4}$. 
To compare it with the magnetic field contribution, we show the expected signal
in Figure~2 for a normalization $B_0 =  1.5 \times 10^{-8} \, \rm G$ for two
values of magnetic spectral indices. 
For theoretical errors on correlation function, the SDSS galaxy survey
might be able to extract the magnetic field contribution for $r \la 30 \, \rm h^{-1} \, Mpc$.   

As seen in Figure~2, changing the value of magnetic field spectral index $n$ doesn't change
our conclusions much. For smaller value of $n$ (theoretically $n > -3$)
there is more power at larger scales. However, qualitatively the results
don't change.

\subsection{Comparison with other constraints}

Several methods have been applied to determine the strength
intergalactic  magnetic fields at $z \la 3$. The most direct method is 
to study the Faraday rotation of the polarized radiation of 
the extra-galactic sources.  For tangled magnetic fields the average
rotation measure (RM) along any line of sight will be zero but it
has non-zero RMS that can be used to constrain the intergalactic magnetic 
field  (for a derivation see Appendix B). The current upper limit on
on RM fluctuations is: $ \la 2 \, {\rm rad \, m^{-2}}$ for $z \simeq 3$ (Vall\'ee 1990).
This gives an upper bound: $B_0 \la 2\hbox{--}3 \times 10^{-8} \, \rm G$ (Appendix B).   The magnetic field power spectrum can
be constrained by correlating the rotation measure 
of extra-galactic radio sources (Kolatt 1998). Using this approach 
it was shown that a few hundred radio sources can be
used to constraint the  magnetic field strengths $\la 10^{-9} \, \rm G$  at scales $10\hbox{--}50 \, \rm Mpc$  (Kollat 1998). To compare it 
with the magnetic field strengths used in this paper we show in Figure~4 the 
Magnetic field RMS smoothed over different scales. For $B_0 = 
1.5 \times 10^{-8}$ this method can be used to constrain magnetic fields
for scale  $\ga 10 \, \rm Mpc$.

Blasi, Burles and Olinto (1999) obtain bounds on magnetic field 
strengths at scales from Jeans' to Horizon scale assuming Lyman-$\alpha$
clouds to trace the matter inhomogeneities and that the magnetic fields
are frozen in the plasma. They showed that magnetic field RMS smoothed
at different scales $\la 10^{-8} \, \rm G$ at these scales. 

All these constraints rule out $B_0 \ga 3 \times 10^{-8}$ which 
is also ruled out by the current galaxy data.  Our requirement
that $B_0 \ga  10^{-8}$ for the magnetic field to leave detectable
signal at linear scales is marginally acceptable by these bounds. 

Subramanian and Barrow (1998,2002) calculated the level of CMBR anisotropies
if the tangled magnetic fields existed at the last scattering surface.
 They concluded that a magnetic field 
strength $B_0 \simeq 3 \times 10^{-9} \, \rm G$ can cause temperature 
fluctuations at the level $\simeq 10 \, \rm \mu K$ for $\ell \simeq 1000\hbox{--}3000$. This is comparable with the detected level of anisotropies 
at these scales (Mason {\it et al.} 2002). Therefore the magnetic field
required to make appreciate effect on the two-point correlation function
could not have existed at the last scattering surface. 

\section{Conclusions}

In this paper we investigated the role of primordial tangled magnetic fields
in shaping the large scale structure in the universe at present. In particular
we calculated the two-point correlation function in the redshift space in the presence of these
fields  for linear scales $\ga 10 \, \rm h^{-1} \, Mpc$ at the present epoch. Our results can be 
summarized as:
\begin{itemize}
\item[1.] Magnetic field contribution to the clustering in redshift space 
is comparable to the gravity-driven clustering for magnetic field 
strength $B_0 \simeq 5 \times 10^{-8}$. 
\item[2.] Present data might have shown the presence of tangled 
 magnetic fields in the two-point correlation function 
for $B_0 \simeq 3 \times 10^{-8}$.
\item[3.] Comparing this with the other bounds on the magnetic field strength
at the present epoch such large magnetic field are mostly likely already 
ruled out. Therefore present data is consistent with this requirement. 
\item[4.] On-going galaxy surveys like SDSS can probe magnetic field 
$B_0 \simeq  10^{-8}$. 
\item[5.] $B_0 \simeq  10^{-8}$ is still too large to be compatible with the CMBR anisotropy
constraints on the primordial magnetic fields. Therefore if the magnetic field
signature is ever detected in the two-point correlation function, it would 
imply that these magnetic fields originated in the post-recombination era.
\end{itemize}

\section*{Acknowledgments}
I would like to thank John Barrow, Ofer Lahav, Martin Rees, Tarun Saini, and 
Kandaswamy Subramanian for useful discussions and suggestions. I would also
like to thank the hospitality of  Institute of Astronomy, Cambridge, UK, where a 
part of this work was done.
 
\section*{Appendix A}
We start with Eq.~(\ref{eq:n4p}). The Green function for the 
homogeneous part can be written as:
\begin{equation}
G^+(t,t') = c(t') u_1(t) + d(t') u_2(t)
\end{equation}
with
\begin{eqnarray}
c(t) & = &  {1 \over a^2(t)} {u_2(t) \over (\dot u_1(t)u_2(t) - u_1(t)\dot u_2(t))} \\
d(t) & = &  {1 \over a^2(t)} {u_1(t) \over (\dot u_1(t)u_2(t) - u_1(t)\dot u_2(t))}
\end{eqnarray}
Here $u_1(t)$ and $u_2(t)$ are the growing and decaying solutions 
of the left hand side of  Eq.~(\ref{eq:n4p}). For a spatially flat 
universe with non-zero cosmological constant:
\begin{eqnarray}
u_1(t) & = & H \int_{0}^a {da' \over H^3 a^3} \\
u_2(t) & =  & H
\end{eqnarray}
Here $H$ the expansion rate is:
\begin{equation}
H^2 = H_0^2 \left (\Omega_m (1+z)^3 + \Omega_\Lambda \right )
\end{equation}
The solution of Eq.~(\ref{eq:n4p}) can be written as:
\begin{equation}
\delta(x,t) = -\int dt' G^+(t,t'){{\bf \nabla .}\left [\left({\bf \nabla} \times {\bf B}(x,t') \right ) \times  {\bf B}(x,t') \right ] \over 4 \pi \rho_{\scriptscriptstyle b}(t')}
\end{equation}
Using  Eq.~(\ref{eq:n5}) the time dependent part of  $\delta(x,t)$ can be 
written as:
\begin{equation}
g(t) = \int dt' G^+(t,t'){1 \over 4 \pi a^6(t') \rho_{\scriptscriptstyle b}(t')}
\end{equation}
The time dependent part of the velocity field can be got by directly 
integrating Eq.~(\ref{eq:n1}) and using Eq.~(\ref{eq:n5}). For the magnetic 
field as the source of large scale velocities: 
\begin{equation}
q(t) = {1\over a(t)} \int dt' {1 \over 4 \pi a^4(t') \rho_{\scriptscriptstyle b}(t')}
\end{equation}
The first term on the right hand side of  Eq.~(\ref{eq:n1}) gives the 
usual gravitational instability growth. 

\section*{Appendix B}
The rotation measure of a source at a redshift $z$ is:
\begin{equation}
RM = 8.1 {\rm rad \, m^{-2}}\left ( {n_e(t_0) \over 10^{-5} \, {\rm cm^{-3}} } \right ) \left ({H_0^{-1} \over 1 \, {\rm Mpc} } \right ) \int_0^z dz'{(1+z')^2 {\bf  \hat z.B}({\bf r})\over (\Omega_m (1+z')^3 + \Omega_\Lambda)^{1/2}} 
\end{equation}
Here $n_e(t_0)$ is the present (ionized) electron density, ${\bf  \hat z.B}({\bf r})$
is the component of magnetic field along the line of sight 
at the present epoch. As the magnetic field is tangled, the line of sight
component has zero mean which implies, $\langle RM \rangle = 0$. The 
average RMS of the rotation measure is:
 \begin{eqnarray}
\langle (RM)^2 \rangle^{1/2} & = & 8.1 {\rm rad \, m^{-2}}\left ( {n_e(t_0) \over 10^{-5} \, {\rm cm^{-3}} } \right ) \left ({H_0^{-1} \over 1 \, {\rm Mpc} } \right ) \times \nonumber \\
& & \!\!\!\!\!\!\!\!\!\!\!\!\!\!\! \left ( \int_0^z dz'' \int_0^z dz'{(1+z')^2 (1+z'')^2\langle {\bf  \hat z.B}({\bf r'}){\bf  \hat z.B}({\bf r''}) \rangle \over (\Omega_m (1+z')^3 + \Omega_\Lambda)^{1/2}(\Omega_m (1+z'')^3 + \Omega_\Lambda)^{1/2}} \right )^{1/2} \nonumber 
\end{eqnarray}
Using Eq.~(\ref{eq:n6}) this expression can be simplified to:
 \begin{eqnarray}
\langle (RM)^2 \rangle^{1/2} & = & 8.1 {\rm rad \, m^{-2}}\left ( {n_e(t_0) \over 10^{-5} \, {\rm cm^{-3}} } \right ) \left ({H_0^{-1} \over 1 \, {\rm Mpc} } \right ) \times \nonumber \\
& & \!\!\!\!\!\!\!\!\!\!\!\!\!\!\! \left(\int_0^z dz'' \int_0^z dz'{(1+z')^2 (1+z'')^2  \over (\Omega_m (1+z')^3 + \Omega_\Lambda)^{1/2}(\Omega_m (1+z'')^3 + \Omega_\Lambda)^{1/2}} \times  \right . \nonumber\\
&& \left .\int {d^3k \over (2\pi)^3} (1-\mu^2)\exp[i\mu k(r'-r'')] B^2(k) \right )^{1/2}
\end{eqnarray}
Here $\mu = {\bf z.k}$. Carrying out angular integral in the k-space
one obtains:
\begin{eqnarray}
\langle (RM)^2 \rangle^{1/2} & = & 8.1 {\rm rad \, m^{-2}}\left ( {n_e(t_0) \over 10^{-5} \, {\rm cm^{-3}} } \right ) \left ({H_0^{-1} \over 1 \, {\rm Mpc} } \right ) \times \nonumber \\
& & \!\!\!\!\!\!\!\!\!\!\!\!\!\!\! \left(\int_0^z dz'' \int_0^z dz'{(1+z')^2 (1+z'')^2  \over (\Omega_m (1+z')^3 + \Omega_\Lambda)^{1/2}(\Omega_m (1+z'')^3 + \Omega_\Lambda)^{1/2}} \times  \right . \nonumber\\
&& \left .\int {dk k^2 \over \pi^2} {j_1\left(k(r'-r'')\right ) \over k(r'-r'')}  B^2(k) \right )^{1/2}
\end{eqnarray}
This expression can be further simplified by using the fact that it
falls rapidly as   $r \equiv r' -r''$ increases, so most 
of the contribution comes from small $r$. This allows one to 
write $r \simeq H^{-1}(z) (z'-z'')$. Making a further change in variable one
gets:
\begin{eqnarray}
\langle (RM)^2 \rangle^{1/2} & \simeq  & 8.1 {\rm rad \, m^{-2}}\left ( {n_e(t_0) \over 10^{-5} \, {\rm cm^{-3}} } \right ) \left ({H_0^{-1} \over 1 \, {\rm Mpc} } \right ) \times \nonumber \\
& & \!\!\!\!\!\!\!\!\!\!\!\!\!\!\! \left( \int_0^z dz'{(1+z')^4   \over (\Omega_m (1+z')^3 + \Omega_\Lambda)^{1/2}} \times  \right . \nonumber\\
&& \left . H_0 \int_0^{r_{\rm max}} dr \int {dk k^2 \over \pi^2} {j_1\left(kr\right ) \over kr}  B^2(k) \right )^{1/2}
\label{eq:a24}
\end{eqnarray} 
For $\Omega_b h^2 =0.02$ one gets $n_e(t_0) = 2.3 \times 10^{-7} \, \rm cm^{-3}$ for a fully ionized universe. 
 For magnetic spectral index $n =1$ and $z = 3$, Eq.~(\ref{eq:a24}) can be numerically solved to give:
\begin{equation}
\langle (RM)^2 \rangle^{1/2} = 0.7 {\rm rad \, m^{-2}} h^{-1/2} \left ( B_0 \over 10^{-8} \, {\rm G} \right) 
\end{equation}
 For magnetic spectral index $n =-2$, the normalization changes from $0.7$ to
$1.2$.

\begin{figure}
\epsfig{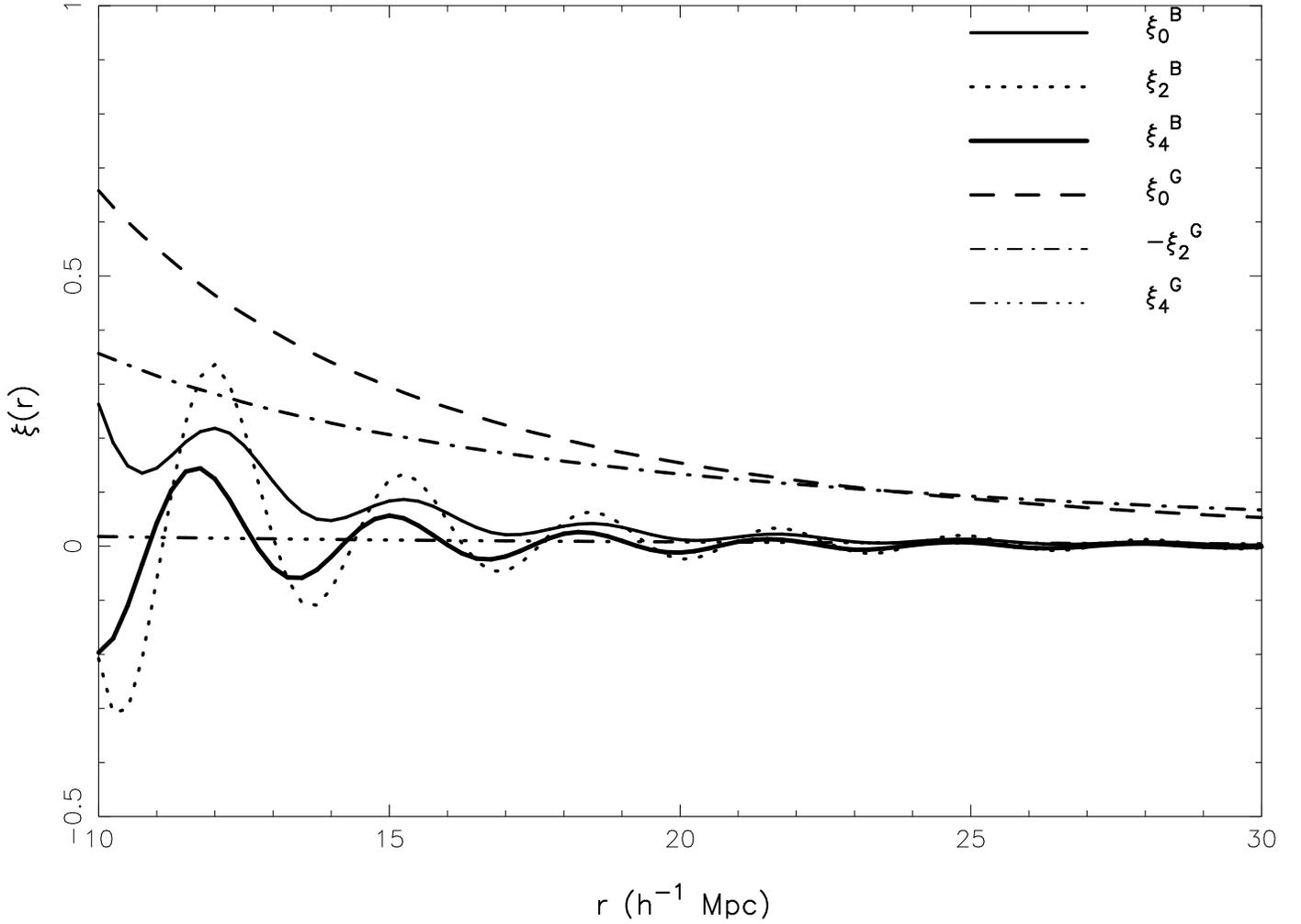}
\caption{The moments of two-point correlation function in redshift space 
are plotted for $B_0 = 5 \times 10^{-8} \, \rm G$ (with superscript 'B')
to indicate magnetic field). Also shown are the three moments for linear 
gravitational
clustering (superscript 'G') for $\beta = 0.43$ (Peacock {\it et al.} 2001)
}
\label{fig:f1}
\end{figure}

\newpage

\begin{figure}
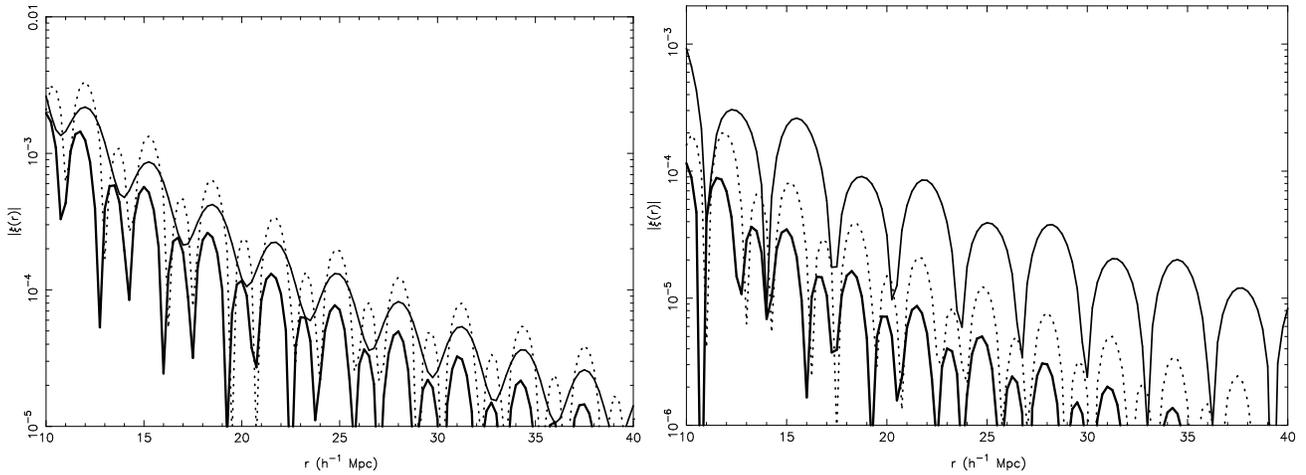

\epsfig{file=fig2_mag.ps,width=0.35\textwidth,angle=270}
\epsfig{file=fig3_mag.ps,width=0.35\textwidth,angle=270}
\caption{The absolute value of the moments of the two-point correlation function is plotted for two values of magnetic spectral index for $B_0 = 1.5 \times 10^{-8} \, \rm G$  .
 The line styles for different moments  is the same as Figure~\ref{fig:f1}. Left Panel: magnetic
spectral index $n = 1$, Right Panel: magnetic spectral index $n = -2$
}
\label{fig:f2}
\end{figure}

\newpage

\begin{figure}
\epsfig{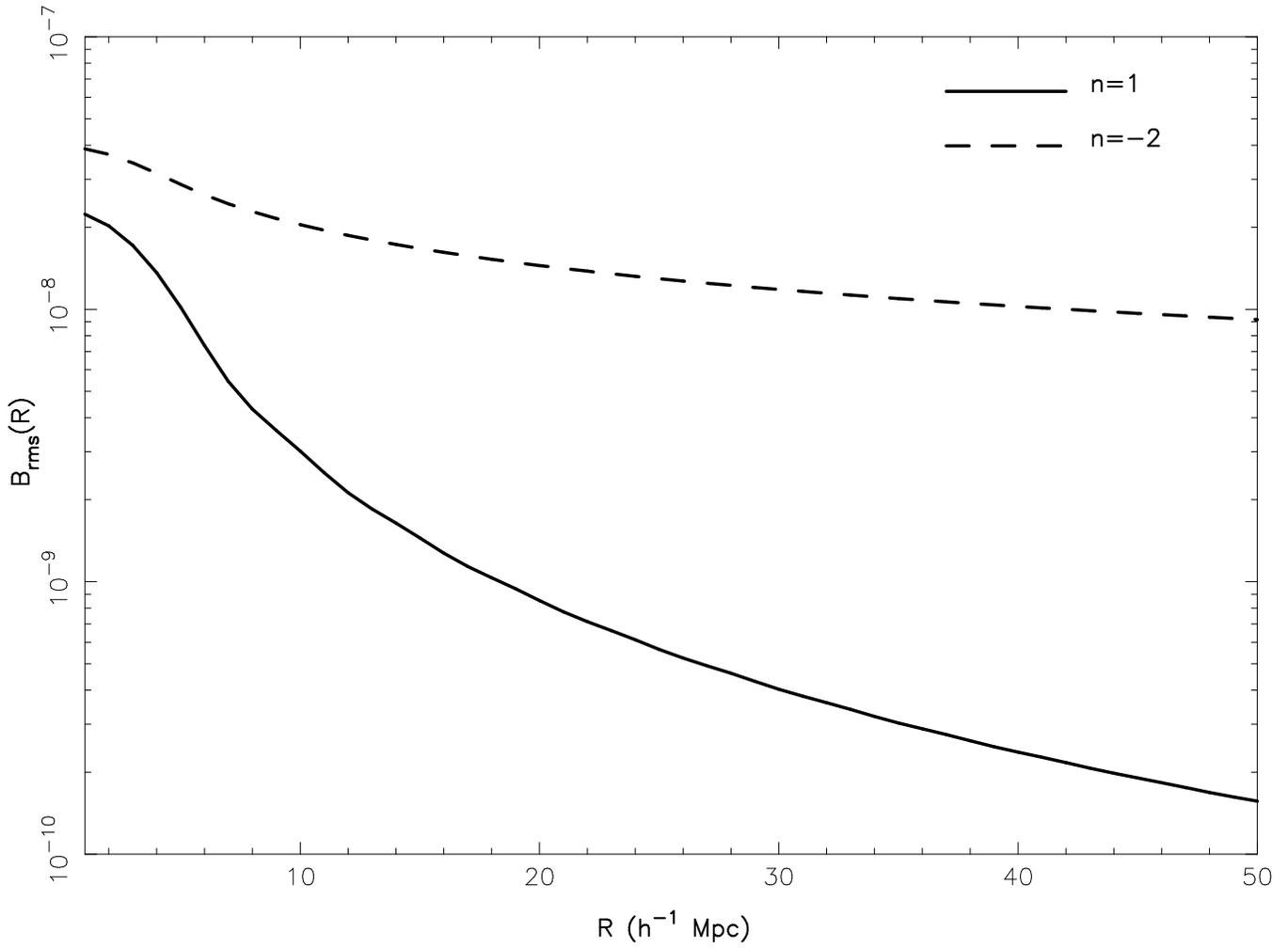}
\caption{The RMS of the smoothed magenetic field fluctuations is plotted as
a function of the smoothing scale for two values of magnetic spectral index
for $B_0 = 1.5 \times 10^{-8} \, \rm G$ }
\label{fig:f3}
\end{figure}


\begin{thebibliography}{}
\bibitem[Bardeen, Bond, Kaiser, \& Szalay(1986)]{1986ApJ...304...15B} 
Bardeen, J.~M., Bond, J.~R., Kaiser, N., \& Szalay, A.~S.\ 1986, \apj, 304, 
15 
\bibitem[Barrow, Ferreira, \& Silk(1997)]{1997PhRvL..78.3610B} Barrow, 
J.~D., Ferreira, P.~G., \& Silk, J.\ 1997, Physical Review Letters, Volume 
78, Issue 19, May 12, 1997, pp.3610-3613, 78, 3610 
\bibitem[Blasi, Burles, \& Olinto(1999)]{1999ApJ...514L..79B} Blasi, P., 
Burles, S., \& Olinto, A.~V.\ 1999, \apjl, 514, L79
\bibitem[Colless et al.(2001)]{2001MNRAS.328.1039C} Colless, M.~et al.\ 
2001, \mnras, 328, 1039
\bibitem[Freedman et al.(2001)]{2001ApJ...553...47F} Freedman, W.~L.~et 
al.\ 2001, \apj, 553, 47 
 \bibitem[Hamilton(1998)]{1998evun.work..185H} Hamilton, A.~J.~S.\ 1998, 
ASSL Vol.~231: The Evolving Universe, 185
\bibitem[Hamilton(1992)]{1992ApJ...385L...5H} Hamilton, A.~J.~S.\ 1992, 
\apjl, 385, L5 
\bibitem[]{} Hawkins, E.~et al.\ 2002, astro-ph/0212375
\bibitem[Hatton \& Cole(1998)]{1998MNRAS.296...10H} Hatton, S.~\& Cole, S.\ 
1998, \mnras, 296, 10
\bibitem[Jedamzik, Katalini{\' c}, \& Olinto(2000)]{2000PhRvL..85..700J} 
Jedamzik, K., Katalini{\' c}, V., \& Olinto, A.~V.\ 2000, Physical Review 
Letters, Volume 85, Issue 4, July 24, 2000, pp.700-703, 85, 700
\bibitem[Kaiser(1987)]{1987MNRAS.227....1K} Kaiser, N.\ 1987, \mnras, 227, 
1 
\bibitem[Kim, Olinto, \& Rosner(1996)]{1996ApJ...468...28K} Kim, E., 
Olinto, A.~V., \& Rosner, R.\ 1996, \apj, 468, 28
\bibitem[Kolatt(1998)]{1998ApJ...495..564K} Kolatt, T.\ 1998, \apj, 495, 
564 
\bibitem[Kosowsky \& Loeb(1996)]{1996ApJ...469....1K} Kosowsky, A.~\& Loeb, 
A.\ 1996, \apj, 469, 1 
\bibitem[Kronberg(1994)]{1994RPPh...57..325K} Kronberg, P.~P.\ 1994, 
Reports on Progress in Physics, Volume 57, Issue 4, pp.~325-382 (1994)., 
57, 325 
\bibitem[Kronberg \& Simard-Normandin(1976)]{1976Natur.263..653K} Kronberg, 
P.~P.~\& Simard-Normandin, M.\ 1976, \nat, 263, 653 
\bibitem[Lahav et al.(2002)]{2002MNRAS.333..961L} Lahav, O.~et al.\ 2002, 
\mnras, 333, 961 
\bibitem[]{} Landau, L. D.  \& Lifshitz, E. M. 1987, Fluid Mechanics, Pergamon Press
\bibitem[Landy \& Szalay(1993)]{1993ApJ...412...64L} Landy, S.~D.~\& 
Szalay, A.~S.\ 1993, \apj, 412, 64
\bibitem[]{} Mason, B. S. ~et al.\ 2002, astro-ph/0205384
\bibitem[Peebles(1980)]{1980lssu.book.....P} Peebles, P.~J.~E.\ 1980, 
Large Scale Structure of the Universe,
Princeton University Press,~435 p.,
\bibitem[Peacock et al.(2001)]{2001Natur.410..169P} Peacock, J.~A.~et al.\ 
2001, \nat, 410, 169 
\bibitem[Perlmutter et al.(1999)]{1999ApJ...517..565P} Perlmutter, S.~et 
al.\ 1999, \apj, 517, 565
\bibitem[Rees \& Reinhardt(1972)]{1972A&A....19..189R} Rees, M.~J.~\& 
Reinhardt, M.\ 1972, \aap, 19, 189 
\bibitem[Riess et al.(1998)]{1998AJ....116.1009R} Riess, A.~G.~et al.\ 
1998, \aj, 116, 1009 
\bibitem[Subramanian \& Barrow(2002)]{2002MNRAS.335L..57S} Subramanian, 
K.~\& Barrow, J.~D.\ 2002, \mnras, 335, L57 
\bibitem[Subramanian \& Barrow(1998)]{1998PhRvL..81.3575S} Subramanian, 
K.~\& Barrow, J.~D.\ 1998, Physical Review Letters, Volume 81, Issue 17, 
October 26, 1998, pp.3575-3578, 81, 3575 
\bibitem[Tytler, O'Meara, Suzuki, \& Lubin(2000)]{2000PhR...333..409T} 
Tytler, D., O'Meara, J.~M., Suzuki, N., \& Lubin, D.\ 2000, \physrep, 333, 
409 
\bibitem[Vallee(1990)]{1990ApJ...360....1V} Vall\'ee, J.~P.\ 1990, \apj, 360, 
1 
\bibitem[Verde et al.(2002)]{2002MNRAS.335..432V} Verde, L.~et al.\ 2002, 
\mnras, 335, 432 
\bibitem[Wasserman(1978)]{1978ApJ...224..337W} Wasserman, I.\ 1978, \apj, 
224, 337
\bibitem[Widrow(2002)]{2002RvMP...74..775W} Widrow, L.~M.\ 2002, Reviews of 
Modern Physics, vol.~74, Issue 3, pp.~775-823, 74, 775 
\bibitem[York et al.(2000)]{2000AJ....120.1579Y} York, D.~G.~et al.\ 2000, 
\aj, 120, 1579 


\newpage

\end{thebibliography}
\end{document}